\documentclass[10pt]{article}
\usepackage{graphicx}
\usepackage{amsmath}
\usepackage{amssymb}
\usepackage{caption2}
\begin{document}
\begin{center}
{\large {\bf \sc{  Analysis of the $D_{s3}^*(2860)$ as a D-wave $c\bar{s}$  meson with  QCD sum rules }}} \\[2mm]
Zhi-Gang Wang \footnote{E-mail,zgwang@aliyun.com.  }      \\
 Department of Physics, North China Electric Power University,
Baoding 071003, P. R. China
\end{center}

\begin{abstract}
In this article, we assign the $D_{s3}^*(2860)$ to be  a D-wave $c\bar{s}$  meson, and  study the mass and decay constant of the $D_{s3}^*(2860)$ with the QCD sum rules by calculating the contributions of the vacuum condensates up to dimension-6 in the operator product expansion. The predicted mass $M_{D_{s3}^*}=(2.86\pm0.10)\,\rm{GeV}$ is in excellent agreement with the experimental value $M_{D_{s3}^*}=(2860.5\pm 2.6 \pm 2.5\pm 6.0)\,\rm{ MeV}$ from the LHCb collaboration. The present prediction supports  assigning the $D_{s3}^*(2860)$ to be the D-wave $c\bar{s}$ meson.
\end{abstract}



\section{Introduction}

In 2006, the BaBar  collaboration  observed the $D^*_{sJ}(2860)$ meson with the mass $(2856.6 \pm 1.5  \pm 5.0)\, \rm{ MeV}$ and the width $(48 \pm 7 \pm 10)\, \rm{ MeV}$ in decays  to the final states  $D^0 K^+$ and $D^+K^0_S$ using $240\rm{ fb}^{-1}$ of data recorded by the BaBar detector at the PEP-II asymmetric-energy $e^+e^-$ storage rings at the Stanford Linear Accelerator Center \cite{BaBar2006}.
In 2009, the BaBar  collaboration confirmed the $D^*_{sJ}(2860)$  in the $D^*K$ channel using $470 {\rm fb}^{-1}$ of data recorded by the BaBar detector, and measured the ratio $R$ among the branching fractions \cite{BaBar2009},
 \begin{eqnarray}
R&=& \frac{{\rm Br}\left(D_{sJ}^*(2860)\to D^*K\right)}{{\rm Br}\left(D_{sJ}^*(2860)\to D K\right)}=1.10 \pm 0.15 \pm 0.19\, \, .
 \end{eqnarray}
The observation of the  decays $D^*_{sJ}(2860)\to D^*K$ rules out the $J^P=0^+$ assignment, the possible assignments are
   the $1^3{\rm D}_3$ $c\bar{s}$ meson \cite{Colangelo0607,Zhang2007,Li2007,Zhong2008,Chen2009,Zhong2010,Li0911,Badalian2011}, the $c\bar{s}-cn\bar{s}\bar{n}$ mixing state \cite{Vijande2009}, the dynamically generated  $D_1(2420)K$ bound state \cite{FKGuo}, etc.

In 2014,  the LHCb collaboration  observed a structure at $2.86\,\rm{GeV}$ with significance of more than $10 \sigma$ in the $\overline{D}^0K^-$ mass spectrum in the Dalitz plot analysis of the decays $B_s^0\to \overline{D}^0K^-\pi^+$ using a data sample
corresponding to an integrated luminosity of $3.0\rm{ fb}^{-1}$ of $pp$ collision data recorded
by the LHCb detector,  the structure contains both spin-1 ($D_{s1}^{*-}(2860)$) and spin-3 ($D_{s3}^{*-}(2860)$) components, which
  can be assigned to be the $J^P =1^-$ and $3^-$ members of the 1D family \cite{LHCb7574,LHCb7712}. The measured masses and widths are
$M_{D_{s3}^*}=(2860.5\pm 2.6 \pm 2.5\pm 6.0)\,\rm{ MeV}$, $M_{D_{s1}^*}=(2859 \pm 12 \pm 6 \pm 23)\,\rm{ MeV}$,
$\Gamma_{D_{s3}^*}=(53 \pm 7 \pm 4 \pm 6)\,\rm{ MeV}$, and $\Gamma_{D_{s1}^*}=(159 \pm 23\pm 27 \pm 72)\,\rm{ MeV}$, respectively.
Furthermore, the LHCb collaboration obtained the conclusion that the $D^*_{sJ}(2860)$ observed by the BaBar collaboration in the inclusive $e^+e^- \to \overline{D}^0K^{-}X$ production  and by the LHCb collaboration in the $pp \to \overline{D}^0K^{-}X$ processes  consists of at least two particles  \cite{BaBar2009,LHCb1207}.

If we assign the $D^*_{sJ}(2860)$ to be the $1^3{\rm D}_3$ state or the $D^*_{s3}(2860)$, the ratio $R$ from the leading order
 heavy meson effective theory \cite{Colangelo0607},
 the ${}^3{\rm P}_0$ model \cite{Zhang2007,Li0911,Song2014} and the relativized quark model  \cite{Godfrey2013} cannot reproduce the experimental value $R=1.10 \pm 0.15 \pm 0.19$ \cite{BaBar2009}.
In Ref.\cite{WangEPJC2860}, we assign the $D_{s3}^*(2860)$ and $D_{s1}^*(2860)$ to be  the $1^3{\rm D}_3$ and $1^3{\rm D}_1$ $c\bar{s}$ states, respectively,  study   their  strong decays   with the heavy meson  effective theory by including the chiral symmetry breaking corrections. We can reproduce the experimental value  $R =1.10 \pm 0.15 \pm 0.19$ with suitable hadronic coupling constants if the chiral symmetry breaking corrections are large. The preferred assignment is $D^*_{sJ}(2860)=D^*_{s3}(2860)$, while the assignment $D^*_{sJ}(2860)=D^*_{s1}(2860)$ is not excluded.

 According to the predictions of the potential models \cite{mass-PQM}, the masses of the 1D $c\bar{s}$ states are  about $2.9\,\rm{GeV}$.
     It is reasonable to assign the $D_{s1}^*(2860)$ and   $D_{s3}^*(2860)$  to be the $\rm{1^3D_1}$ and $\rm{1^3D_3}$  $c\bar{s}$ states, respectively. We can obtain further support by calculating the mass of the $D_{s3}^*(2860)$ based on the QCD sum rules.
 The  QCD sum rules is a powerful  theoretical tool in studying the
ground state hadrons and has given many successful descriptions of the masses, decay constants, form-factors and hadronic coupling constants, etc \cite{SVZ79,Reinders85}.
 There have been many works on the spin-parity
$J^P=0^\pm$, $1^\pm$  heavy-light mesons with the full QCD sum rules \cite{WangEPJC-HL,Narison-HL} (and references therein), while the works on the $J^P=2^+$   are few \cite{Azizi-Tmeson,Wang-Tmeson}, the $J^P=3^-$ heavy-light mesons are only studied with the QCD sum rules combined with the heavy quark effective theory \cite{Zhu-D-wave}.
In this article, we assign the $D_{s3}^*(2860)$ to be  a D-wave $c\bar{s}$  meson,  study the mass and decay constant of the  $D_{s3}^*(2860)$ with the full QCD sum rules in details by calculating the contributions of the  vacuum condensates up to dimension-6 in the operator product expansion.

The article is arranged as follows:  we derive the QCD sum rules for
the mass and decay constant of the $D_{s3}^*(2860)$  in Sect.2;
in Sect.3, we present the numerical results and discussions; and Sect.4 is reserved for our
conclusions.

\section{QCD sum rules for  the $D_{s3}^*(2860)$ as a D-wave meson }
In the following, we write down  the two-point correlation function
$\Pi_{\mu\nu\rho\alpha\beta\sigma}(p)$  in the QCD sum rules,
\begin{eqnarray}
\Pi_{\mu\nu\rho\alpha\beta\sigma}(p)&=&i\int d^4x e^{ip \cdot (x-y)} \langle
0|T\left\{J_{\mu\nu\rho}(x)J_{\alpha\beta\sigma}^{\dagger}(y)\right\}|0\rangle\mid_{y=0} \, ,
\end{eqnarray}
where the current
\begin{eqnarray}
J_{\mu\nu\rho}(x)&=&\overline{c}(x)\left( \gamma_\mu\stackrel{\leftrightarrow}{D}_\nu\stackrel{\leftrightarrow}{D}_\rho
+\gamma_\nu\stackrel{\leftrightarrow}{D}_\rho\stackrel{\leftrightarrow}{D}_\mu+\gamma_\rho\stackrel{\leftrightarrow}{D}_\mu\stackrel{\leftrightarrow}{D}_\nu \right) s(x) \, ,
\end{eqnarray}
with $\stackrel{\leftrightarrow}{D}_\mu=\stackrel{\rightarrow}{\partial}_\mu-ig_sG_\mu-\stackrel{\leftarrow}{\partial}_\mu-ig_sG_\mu $
  interpolates the D-wave meson $D_{s3}^*(2860)$. We can rewrite the current into two parts,
\begin{eqnarray}
J_{\mu\nu\rho}(x)&=&\eta_{\mu\nu\rho}(x)+J^V_{\mu\nu\rho}(x)\, ,
\end{eqnarray}
where
\begin{eqnarray}
\eta_{\mu\nu\rho}(x)&=&\overline{c}(x)\left( \gamma_\mu\stackrel{\leftrightarrow}{\partial}_\nu\stackrel{\leftrightarrow}{\partial}_\rho
+\gamma_\nu\stackrel{\leftrightarrow}{\partial}_\rho\stackrel{\leftrightarrow}{\partial}_\mu+\gamma_\rho\stackrel{\leftrightarrow}{\partial}_\mu\stackrel{\leftrightarrow}{\partial}_\nu \right) s(x) \, , \\
J^V_{\mu\nu\rho}(x)&=&-2i\,\overline{c}(x)\left[ \gamma_\mu\left( g_sG_\nu\stackrel{\leftrightarrow}{\partial}_\rho
+\stackrel{\leftrightarrow}{\partial}_\nu g_sG_\rho-2ig_s^2 G_\nu G_\rho\right)\right.\nonumber\\
&&\left.+ \gamma_\nu\left( g_sG_\rho\stackrel{\leftrightarrow}{\partial}_\mu
+\stackrel{\leftrightarrow}{\partial}_\rho g_sG_\mu-2ig_s^2 G_\rho G_\mu\right)\right.\nonumber\\
&&\left.+\gamma_\rho\left( g_sG_\mu\stackrel{\leftrightarrow}{\partial}_\nu
+\stackrel{\leftrightarrow}{\partial}_\mu g_sG_\nu-2ig_s^2 G_\mu G_\nu\right)\right] s(x) \, ,
\end{eqnarray}
with $\stackrel{\leftrightarrow}{\partial}_\mu=\stackrel{\rightarrow}{\partial}_\mu-\stackrel{\leftarrow}{\partial}_\mu$ and the $G_\mu$ is the gluon field.

 We  can choose either the partial  derivative $\partial_\mu$ or the covariant derivative $D_\mu$ to construct  the interpolating  currents. The current $J_{\mu\nu\rho}(x)$ with the covariant derivative $D_\mu$ is gauge invariant, but blurs the physical interpretation of the $\stackrel{\leftrightarrow}{D}_\mu$ being the angular momentum. The current $\eta_{\mu\nu\rho}(x)$ with the partial derivative $\partial_\mu$ is not gauge invariant, but manifests the physical interpretation of the $\stackrel{\leftrightarrow}{\partial}_\mu$ being the angular momentum. In this article, we will present the results come from the currents with both the partial derivative and the covariant derivative.

We can insert  a complete set of intermediate hadronic states with
the same quantum numbers as the current operator $J_{\mu\nu\rho}(x)$ into the
correlation function $\Pi_{\mu\nu\rho\alpha\beta\sigma}(p)$  to obtain the hadronic representation
\cite{SVZ79,Reinders85}.
The current $J_{\mu\nu\rho}(0)$ has negative parity, and    couples potentially to the $J^P={3}^-$  $\bar{c}s$ meson $D_{s3}^*(2860)$.
On the other hand, the current $J_{\mu\nu\rho}(0)$  also couples potentially to the $J^P={2}^+$, $1^-$, $0^+$   $\bar{c}s$ mesons,
 \begin{eqnarray}
 \langle 0|J_{\mu\nu\rho}(0)|D_{s3}^*(p)\rangle&=&f_{D_{s3}^*}\varepsilon_{\mu\nu\rho}(p,\lambda)  \, ,\\
\langle 0| J_{\mu\nu\rho}(0)|D_{s2}^*(p)\rangle &=&f_{D_{s2}^*} \left[ p_\mu \varepsilon_{\nu\rho}(p,\lambda)+p_\nu\varepsilon_{\rho\mu}(p,\lambda)+p_\rho\varepsilon_{\mu\nu}(p,\lambda)\right] \, , \nonumber\\
\langle 0| J_{\mu\nu\rho}(0)|D_{s1}^*(p)\rangle &=&f_{D_{s1}^*} \left[ p_\mu p_\nu \varepsilon_{\rho}(p,\lambda)+p_\nu p_\rho \varepsilon_{\mu}(p,\lambda)+p_\rho p_\mu\varepsilon_{\nu}(p,\lambda)\right] \, , \nonumber\\
\langle 0| J_{\mu\nu\rho}(0)|D_{s0}^*(p)\rangle &=&f_{D_{s0}^*}  p_\mu p_\nu p_{\rho} \, ,
\end{eqnarray}
where the $f_{D_{s3}^*}$, $f_{D_{s2}^*}$, $f_{D_{s1}^*}$ and $f_{D_{s0}^*}$ are the decay constants, the $\varepsilon_{\mu\nu\rho}(p,\lambda)$, $\varepsilon_{\mu\nu}(p,\lambda)$ and $\varepsilon_{\mu}(p,\lambda)$ are the polarization vectors of the $\bar{c}s$ mesons  with the following properties \cite{JJZhu},
\begin{eqnarray}
{\rm P}_{\mu\nu\rho\alpha\beta\sigma}&=&\sum_{\lambda} \varepsilon^*_{\mu\nu\rho}(\lambda,p)\varepsilon_{\alpha\beta\sigma}(\lambda,p)  \nonumber\\
&=&\frac{1}{6}\left(\widetilde{g}_{\mu\alpha}\widetilde{g}_{\nu\beta}\widetilde{g}_{\rho\sigma}+\widetilde{g}_{\mu\alpha}\widetilde{g}_{\nu\sigma}\widetilde{g}_{\rho\beta}
+\widetilde{g}_{\mu\beta}\widetilde{g}_{\nu\alpha}\widetilde{g}_{\rho\sigma}   +\widetilde{g}_{\mu\beta}\widetilde{g}_{\nu\sigma}\widetilde{g}_{\rho\alpha}
 +\widetilde{g}_{\mu\sigma}\widetilde{g}_{\nu\alpha}\widetilde{g}_{\rho\beta}+\widetilde{g}_{\mu\sigma}\widetilde{g}_{\nu\beta}\widetilde{g}_{\rho\alpha}\right)\nonumber\\
&&-\frac{1}{15}\left(\widetilde{g}_{\mu\alpha}\widetilde{g}_{\nu\rho}\widetilde{g}_{\beta\sigma}+\widetilde{g}_{\mu\beta}\widetilde{g}_{\nu\rho}\widetilde{g}_{\alpha\sigma}
+\widetilde{g}_{\mu\sigma}\widetilde{g}_{\nu\rho}\widetilde{g}_{\alpha\beta}  +\widetilde{g}_{\nu\alpha}\widetilde{g}_{\mu\rho}\widetilde{g}_{\beta\sigma}
 +\widetilde{g}_{\nu\beta}\widetilde{g}_{\mu\rho}\widetilde{g}_{\alpha\sigma}   +\widetilde{g}_{\nu\sigma}\widetilde{g}_{\mu\rho}\widetilde{g}_{\alpha\beta}\right. \nonumber\\
&&\left. +\widetilde{g}_{\rho\alpha}\widetilde{g}_{\mu\nu}\widetilde{g}_{\beta\sigma}  +\widetilde{g}_{\rho\beta}\widetilde{g}_{\mu\nu}\widetilde{g}_{\alpha\sigma}
         +\widetilde{g}_{\rho\sigma}\widetilde{g}_{\mu\nu}\widetilde{g}_{\alpha\beta}\right) \, , \\
 {\rm P}_{\mu\nu\alpha\beta}&=& \sum_\lambda \varepsilon^*_{\mu\nu}(\lambda,p)\varepsilon_{\alpha\beta}(\lambda,p)=\frac{\widetilde{g}_{\mu\alpha}\widetilde{g}_{\nu\beta}
 +\widetilde{g}_{\mu\beta}\widetilde{g}_{\nu\alpha}}{2}-\frac{\widetilde{g}_{\mu\nu}\widetilde{g}_{\alpha\beta}}{3}  \, , \\
 \widetilde{g}_{\mu\nu} &=&\sum_\lambda \varepsilon^*_{\mu}(\lambda,p)\varepsilon_{\nu}(\lambda,p)=-g_{\mu\nu}+\frac{p_\mu p_\nu}{p^2}  \, .
  \end{eqnarray}

The correlation function can be written into the following form according to Lorentz covariance,
\begin{eqnarray}
\Pi_{\mu\nu\rho\alpha\beta\sigma}(p)&=&\Pi(p^2){\rm P}_{\mu\nu\rho\alpha\beta\sigma} +\Pi_2(p^2)\left[ {\rm P}_{\nu\rho\beta\sigma}\,p_\mu p_\alpha + {\rm P}_{\nu\rho\alpha\sigma}\,p_\mu p_\beta + {\rm P}_{\nu\rho\alpha\beta}\,p_\mu p_\sigma + {\rm P}_{\mu\rho\beta\sigma}\,p_\nu p_\alpha\right.\nonumber\\
&&\left.+ {\rm P}_{\mu\rho\alpha\sigma}\,p_\nu p_\beta+ {\rm P}_{\mu\rho\alpha\beta}\,p_\nu p_\sigma+ {\rm P}_{\mu\nu\beta\sigma}\,p_\rho p_\alpha+ {\rm P}_{\mu\nu\alpha\sigma}\,p_\rho p_\beta+ {\rm P}_{\mu\nu\alpha\beta}\,p_\rho p_\sigma\right] \nonumber\\
&&+\Pi_1(p^2)\left[\widetilde{g}_{\mu \alpha}\, p_\nu p_\rho p_\beta p_\sigma+\widetilde{g}_{\mu \beta}\, p_\nu p_\rho p_\alpha p_\sigma
+\widetilde{g}_{\mu \sigma}\, p_\nu p_\rho p_\alpha p_\beta +\widetilde{g}_{\nu \alpha}\, p_\mu p_\rho p_\beta p_\sigma\right.\nonumber\\
&&\left.+\widetilde{g}_{\nu \beta} \,p_\mu p_\rho p_\alpha p_\sigma+\widetilde{g}_{\nu \sigma}\, p_\mu p_\rho p_\alpha p_\beta
+\widetilde{g}_{\rho \alpha} \, p_\mu p_\nu p_\beta p_\sigma+\widetilde{g}_{\rho \beta}\, p_\mu p_\nu p_\alpha p_\sigma
+\widetilde{g}_{\rho \sigma}\, p_\mu p_\nu p_\alpha p_\beta\right] \nonumber\\
&&+\Pi_0(p_2)\, p_\mu p_\nu p_\rho p_\alpha p_\beta p_\sigma \, ,
\end{eqnarray}
the components $\Pi_2(p^2)$, $\Pi_1(p^2)$ and $\Pi_0(p^2)$  come from the contributions of the $J^P=2^+$, $1^-$ and $0^+$ $\bar{c}s$ mesons, respectively.

We isolate the ground state contribution from the $D_{s3}^*(2860)$ and get the following result,
\begin{eqnarray}
\Pi_{\mu\nu\rho\alpha\beta\sigma}(p)&=&\frac{f_{D_{s3}^*}^2}{M_{D_{s3}^*}^2-p^2}{\rm P}_{\mu\nu\rho\alpha\beta\sigma} +\cdots\,  ,\nonumber\\
&=&\Pi(p^2){\rm P}_{\mu\nu\rho\alpha\beta\sigma}+\cdots\,  .
\end{eqnarray}
We can project out the component $\Pi(p^2)$,
\begin{eqnarray}
\Pi(p^2)&=&\frac{1}{7}{\rm P}^{\mu\nu\rho\alpha\beta\sigma}\Pi_{\mu\nu\rho\alpha\beta\sigma}(p) \, ,
\end{eqnarray}
according to the properties,
\begin{eqnarray}
p^\mu{\rm P}_{\mu\nu\rho\alpha\beta\sigma}=p^\nu{\rm P}_{\mu\nu\rho\alpha\beta\sigma}=p^\rho{\rm P}_{\mu\nu\rho\alpha\beta\sigma}=p^\alpha{\rm P}_{\mu\nu\rho\alpha\beta\sigma}=p^\beta{\rm P}_{\mu\nu\rho\alpha\beta\sigma}=p^\sigma{\rm P}_{\mu\nu\rho\alpha\beta\sigma}=0\,.
\end{eqnarray}

Now, we briefly outline  the operator product expansion for the correlation function $\Pi_{\mu\nu\rho\alpha\beta\sigma}(p)$  in perturbative
QCD.  We contract the quark fields in the correlation function $\Pi_{\mu\nu\rho\alpha\beta\sigma}(p)$ with Wick theorem firstly,
\begin{eqnarray}
\Pi(p^2)&=&-\frac{i}{7}{\rm P}^{\mu\nu\rho\alpha\beta\sigma}\int d^4x e^{ip \cdot (x-y)}   Tr\left\{\Gamma_{\mu\nu\rho}^{ik}(x)S_{k l}(x-y)\Gamma_{\alpha\beta\sigma}^{l j}(y) S^c_{ji}(y-x) \right\}\mid_{y=0}\, ,
\end{eqnarray}
where $\Gamma_{\mu\nu\rho}(x)$ and $\Gamma_{\alpha\beta\sigma}(y)$ are the vertexes,
\begin{eqnarray}
S_{ij}(x)&=& \frac{i\delta_{ij}\!\not\!{x}}{ 2\pi^2x^4}
-\frac{\delta_{ij}m_s}{4\pi^2x^2}-\frac{\delta_{ij}}{12}\langle\bar{s}s\rangle +\frac{i\delta_{ij}\!\not\!{x}m_s
\langle\bar{s}s\rangle}{48}-\frac{\delta_{ij}x^2\langle \bar{s}g_s\sigma Gs\rangle}{192}+\frac{i\delta_{ij}x^2\!\not\!{x} m_s\langle \bar{s}g_s\sigma
 Gs\rangle }{1152}\nonumber\\
&& -\frac{ig_s G^{a}_{\alpha\beta}t^a_{ij}(\!\not\!{x}
\sigma^{\alpha\beta}+\sigma^{\alpha\beta} \!\not\!{x})}{32\pi^2x^2}    -\frac{1}{8}\langle\bar{s}_j\sigma^{\mu\nu}s_i \rangle \sigma_{\mu\nu}  +\cdots \, ,
\end{eqnarray}
\begin{eqnarray}
S^c_{ij}(x)&=&\frac{i}{(2\pi)^4}\int d^4k e^{-ik \cdot x} \left\{
\frac{\delta_{ij}}{\!\not\!{k}-m_c}
-\frac{g_sG^n_{\alpha\beta}t^n_{ij}}{4}\frac{\sigma^{\alpha\beta}(\!\not\!{k}+m_c)+(\!\not\!{k}+m_c)
\sigma^{\alpha\beta}}{(k^2-m_c^2)^2}\right.\nonumber\\
&& -\frac{g_s^2 (t^at^b)_{ij} G^a_{\alpha\beta}G^b_{\mu\nu}(f^{\alpha\beta\mu\nu}+f^{\alpha\mu\beta\nu}+f^{\alpha\mu\nu\beta}) }{4(k^2-m_c^2)^5}\nonumber\\
&&\left.+\frac{i\langle g_s^3GGG\rangle}{48}\frac{(\!\not\!{k}+m_c)\left[\!\not\!{k}(k^2-3m_c^2)+2m_c(2k^2-m_c^2) \right](\!\not\!{k}+m_c)}{(k^2-m_c^2)^6}+\cdots\right\}\, ,\nonumber\\
f^{\alpha\beta\mu\nu}&=&(\!\not\!{k}+m_c)\gamma^\alpha(\!\not\!{k}+m_c)\gamma^\beta(\!\not\!{k}+m_c)\gamma^\mu(\!\not\!{k}+m_c)\gamma^\nu(\!\not\!{k}+m_c)\, ,
\end{eqnarray}
 $t^n=\frac{\lambda^n}{2}$, the $\lambda^n$ is the Gell-Mann matrix, the $i$, $j$ are color indexes; then compute  the integrals both in
the coordinate and momentum spaces;  finally obtain the QCD spectral density through dispersion relation,
\begin{eqnarray}
\Pi(p^2)&=&\frac{1}{\pi}\int_{m_c^2}^\infty \frac{{\rm Im}\Pi(s)}{s-p^2}=\int_{m_c^2}^\infty \frac{\rho_{QCD}(s)}{s-p^2}\, .
\end{eqnarray}
In Eq.(17), we retain the term $\langle\bar{s}_j\sigma_{\mu\nu}s_i \rangle$ originates  from the Fierz re-ordering of the $\langle s_i \bar{s}_j\rangle$ to  absorb the gluons  emitted from the heavy quark line to form $\langle\bar{s}_j g_s G^a_{\alpha\beta} t^a_{mn}\sigma_{\mu\nu} s_i \rangle$  to extract the mixed condensate $\langle\bar{s}g_s\sigma G s\rangle$. There are no contributions come from the terms $\langle\bar{s}  s\rangle$ and $\langle\bar{s}g_s\sigma G s\rangle$ due to the projector ${\rm P}_{\mu\nu\rho\alpha\beta\sigma}$. So it is convenient to use the  following $s$ quark propagator,
\begin{eqnarray}
S_{ij}(x)&=&S^c_{ij}(x)|_{m_c \to m_s}\, .
\end{eqnarray}
We take into account all the Feynman diagrams shown explicitly in Figs.1-2, which contribute to the gluon condensate and three-gluon condensate.
 In the Feynman diagrams, we use the solid and dashed lines to represent the light and heavy quark propagators, respectively.
In the fixed point gauge, $G_\mu(x)=\frac{1}{2}x^\theta G_{\theta\mu}(0)+\cdots$ and $G_\alpha(y)=\frac{1}{2}y^\theta G_{\theta\alpha}(0)+\cdots=0$. So in the $\Gamma_{\alpha\beta\sigma}(y)$, we can set $G_\alpha(y)=G_\beta(y)=G_\sigma(y)=0$, there are no gluon lines associated with the vertexes   at the point $y=0$ in Fig.2.

\begin{figure}
 \centering
 \includegraphics[totalheight=6cm,width=14cm]{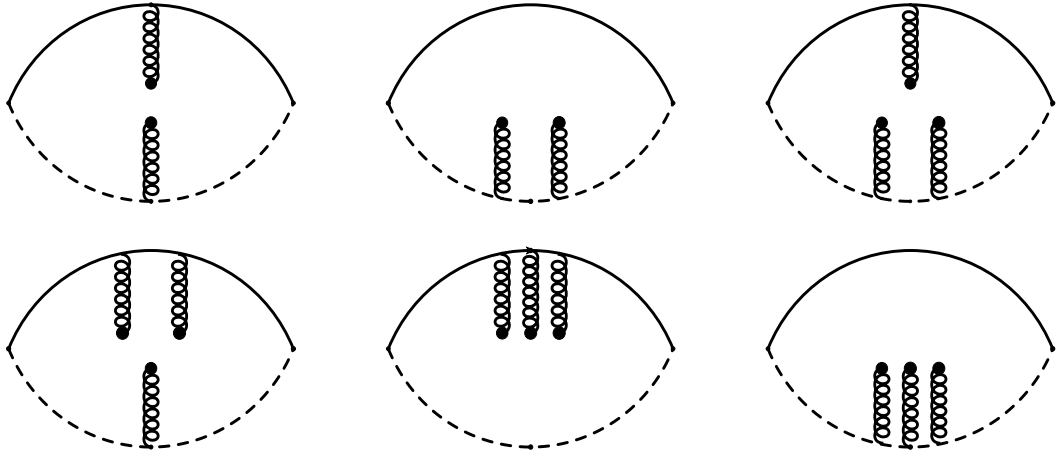}
    \caption{The diagrams contribute to the gluon condensate $\langle \frac{\alpha_sGG}{\pi}\rangle$ and three-gluon condensate $\langle g_s^3 GGG\rangle$. }
\end{figure}

\begin{figure}
 \centering
 \includegraphics[totalheight=3cm,width=14cm]{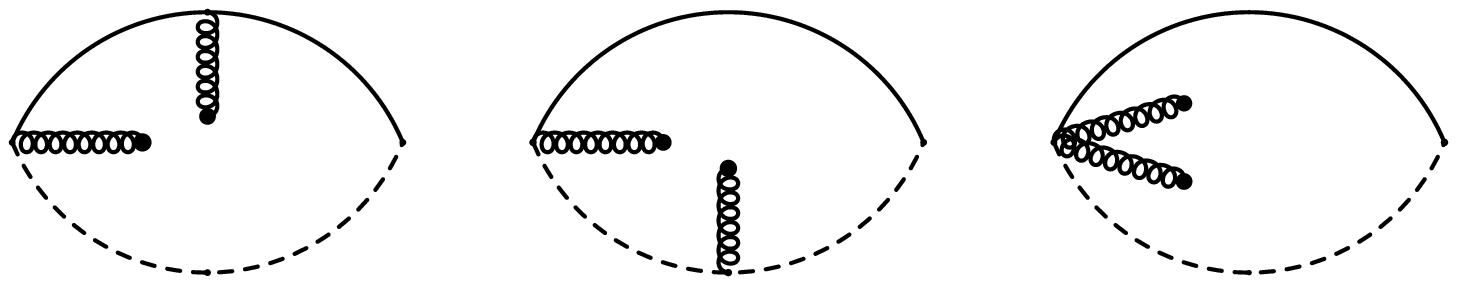}
 \vglue+4mm
 \includegraphics[totalheight=3cm,width=14cm]{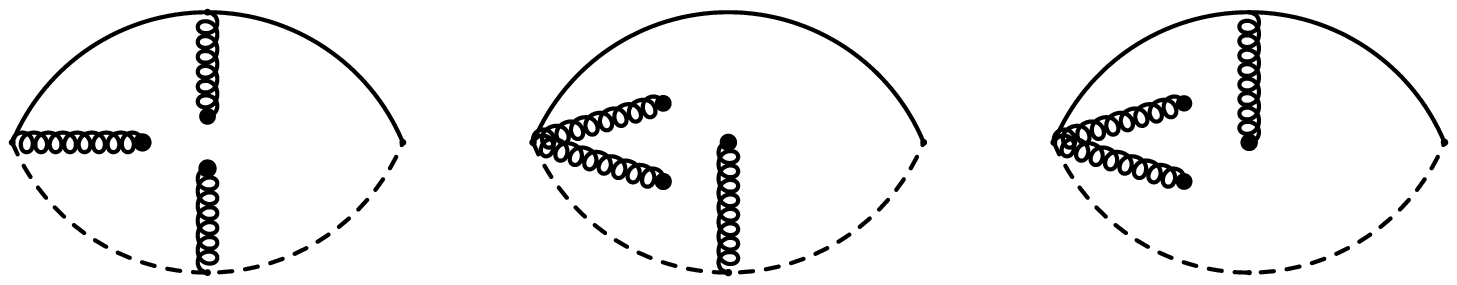}
  \vglue+4mm
 \includegraphics[totalheight=3cm,width=14cm]{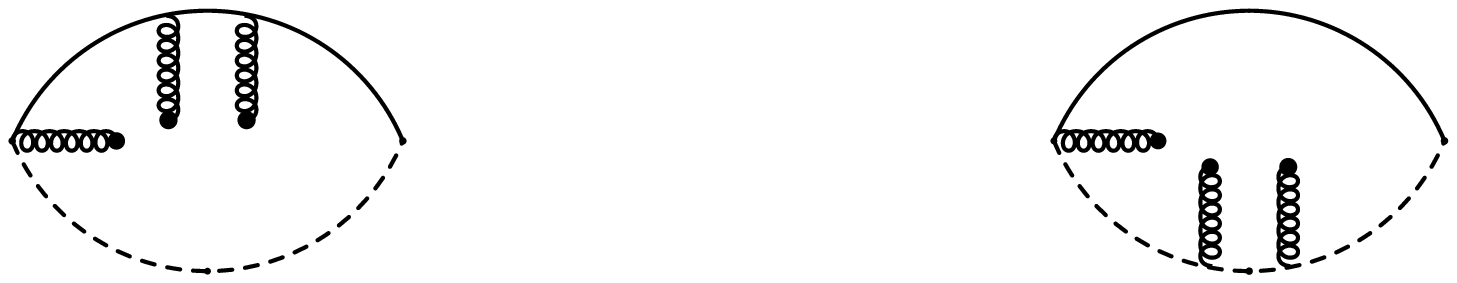}
    \caption{The additional diagrams contribute to the gluon condensate $\langle \frac{\alpha_sGG}{\pi}\rangle$ and three-gluon condensate $\langle g_s^3 GGG\rangle$ from the covariant derivative. }
\end{figure}

We take quark-hadron duality below the continuum threshold $s_0$ and perform the Borel transform  with respect to the variable
$P^2=-p^2$ to obtain the QCD sum rule,
\begin{eqnarray}
 f_{D_{s3}^*}^2 \exp\left(-\frac{M_{D_{s3}^*}^2}{T^2}\right)&=&B_{T^2}\Pi(p^2) \, ,
\end{eqnarray}
where
\begin{eqnarray}
 B_{T^2}\Pi(p^2)&=& \frac{9}{140\pi^2} \int_{m_c^2}^{s_0} ds \frac{(s-m_c^2)^6(4s+3m_c^2)+14m_s m_c s(s-m_c^2)^5}{s^4}  \exp\left(-\frac{s}{T^2}\right) \nonumber \\
  && +m_c^4\langle \frac{\alpha_sGG}{\pi}\rangle  \left\{ \Pi_{GG}^\eta +\Pi_{GG}^V \right\} +\frac{\langle g_s^3 GGG\rangle}{48\pi^2}\left\{ \Pi_{GGG}^\eta +\Pi_{GGG}^V \right\} \, ,
\end{eqnarray}

\begin{eqnarray}
\Pi_{GG}^\eta&=&  \left[\frac{3}{5}+\frac{17m_c^2}{10T^2}+\frac{3m_c^4}{10T^4}+\frac{T^2}{5m_c^2}-\frac{4T^4}{5m_c^4}\right]\exp\left( -\frac{m_c^2}{T^2}\right)- \left[\frac{2m_c^2}{T^2}+\frac{2m_c^4}{T^4}+\frac{3m_c^6}{10T^6}\right]\Gamma\left( 0,\frac{m_c^2}{T^2}\right)\, , \nonumber\\
\end{eqnarray}

\begin{eqnarray}
\Pi_{GGG}^\eta&=& \left[51+\frac{81m_c^2}{T^2}+\frac{12m_c^4}{T^4}-\frac{15T^2}{m_c^2}\right]\exp\left( -\frac{m_c^2}{T^2}\right)- \left[\frac{120m_c^2}{T^2}+\frac{93m_c^4}{T^4}+\frac{12m_c^6}{T^6}\right]\Gamma\left( 0,\frac{m_c^2}{T^2}\right) \nonumber\\
&&+\frac{9m_c^6}{8T^6}\log\frac{m_c^2}{T^2}\, ,
\end{eqnarray}

\begin{eqnarray}
\Pi_{GG}^V&=&  \left[-\frac{11}{10}+\frac{7m_c^2}{40T^2}+\frac{3m_c^4}{40T^4}+\frac{21T^2}{20m_c^2}-\frac{17T^4}{10m_c^4}\right]\exp\left( -\frac{m_c^2}{T^2}\right)+ \left[\frac{m_c^2}{T^2}-\frac{m_c^4}{4T^4}-\frac{3m_c^6}{40T^6}\right]\Gamma\left( 0,\frac{m_c^2}{T^2}\right) \, ,\nonumber\\
\end{eqnarray}

\begin{eqnarray}
\Pi_{GGG}^V&=&  \left[33+\frac{81m_c^2}{2T^2}+\frac{9m_c^4}{2T^4}-\frac{6T^2}{m_c^2}\right]\exp\left( -\frac{m_c^2}{T^2}\right)- \left[\frac{69m_c^2}{T^2}+\frac{45m_c^4}{T^4}+\frac{9m_c^6}{2T^6}\right]\Gamma\left( 0,\frac{m_c^2}{T^2}\right) \, , \nonumber\\
\end{eqnarray}
 $\Gamma(0,x)=\int_0^\infty dt \frac{1}{t}e^{-xt}$, the superscripts $\eta$ and $V$ denote the contributions come from the Feynman diagrams in Fig.1 and Fig.2, respectively.

We differentiate   Eq.(21) with respect to  $\frac{1}{T^2}$, then eliminate the decay constant $f_{D_{s3}^*}$, and obtain the QCD sum rule for  the mass of the $D_{s3}^*(2860)$,
 \begin{eqnarray}
 M_{D_{s3}^*}^2 &=& -\frac{\frac{d }{d(1/T^2)}B_{T^2}\Pi(p^2)}{B_{T^2}\Pi(p^2)} \, .
 \end{eqnarray}
Once the mass $M_{D_{s3}^*}$ is obtained, we can take it as the input parameter and obtain the decay constant from the QCD sum rule in Eq.(21).

\section{Numerical results and discussions}
The values of the gluon condensate and three-gluon condensate can be   taken to be the standard values (SV) $\langle \frac{\alpha_s GG}{\pi}\rangle=0.012 \,\rm{GeV}^4 $ and $\langle g_s^3 GGG\rangle=0.045\,\rm{GeV}^6$ \cite{SVZ79,Reinders85,CReview}.
 The value of the gluon condensate $\langle \frac{\alpha_s
GG}{\pi}\rangle $ has been updated from time to time, and changes
greatly \cite{NarisonBook}, we can choose the recently updated value $\langle \frac{\alpha_s GG}{\pi}\rangle=(0.022 \pm
0.004)\,\rm{GeV}^4 $ \cite{gg-conden}, and take the three-gluon condensate as $\langle g_s^3 GGG\rangle=(8.8\pm5.5)\,{\rm{GeV}^2}\langle\alpha_s GG\rangle=(0.616\pm0.385)\,\rm{GeV}^6$ \cite{gg-conden}. The most recent value of the gluon condensate from the QCD sum rules is $\langle \frac{\alpha_s GG}{\pi}\rangle=(0.037 \pm 0.015)\,\rm{GeV}^4 $ \cite{Dominguez-gg},
but the value of the three-gluon condensate is not updated. Thereafter the recently updated values (UV) of the gluon condensate and three-gluon condensate in Ref.\cite{gg-conden} will be referred to as UV. The SV and UV differ from each other greatly, there are no overlaps between the two sets of parameters,  we obtain the mass and decay constant with the two sets of parameters separately,  one can take the average values.

As the quark masses are concerned, we can
take the $\overline{MS}$ masses $m_{c}(m_c)=(1.275\pm0.025)\,\rm{GeV}$ and $m_s(\mu=2\,\rm{GeV})=(0.095\pm0.005)\,\rm{GeV}$
 from the Particle Data Group \cite{PDG}, and take into account
the energy-scale dependence of  the $\overline{MS}$ masses from the renormalization group equation,
\begin{eqnarray}
m_s(\mu)&=&m_s({\rm 2GeV} )\left[\frac{\alpha_{s}(\mu)}{\alpha_{s}({\rm 2GeV})}\right]^{\frac{4}{9}} \, ,\nonumber\\
m_c(\mu)&=&m_c(m_c)\left[\frac{\alpha_{s}(\mu)}{\alpha_{s}(m_c)}\right]^{\frac{12}{25}} \, ,\nonumber\\
\alpha_s(\mu)&=&\frac{1}{b_0t}\left[1-\frac{b_1}{b_0^2}\frac{\log t}{t} +\frac{b_1^2(\log^2{t}-\log{t}-1)+b_0b_2}{b_0^4t^2}\right]\, ,
\end{eqnarray}
  where $t=\log \frac{\mu^2}{\Lambda^2}$, $b_0=\frac{33-2n_f}{12\pi}$, $b_1=\frac{153-19n_f}{24\pi^2}$, $b_2=\frac{2857-\frac{5033}{9}n_f+\frac{325}{27}n_f^2}{128\pi^3}$,  $\Lambda=213\,\rm{MeV}$, $296\,\rm{MeV}$  and  $339\,\rm{MeV}$ for the flavors  $n_f=5$, $4$ and $3$, respectively  \cite{PDG}. In calculations, we can take $n_f=4$.
 We usually take the energy scale   $\mu_{D}=1\,\rm{GeV}$ for the pseudoscalar $D$ meson, if we count  the contributions of the additional P-wave and D-wave as  $0.5\,\rm{GeV}$ and $1\,\rm{GeV}$, respectively, and assume  the flavor $SU(3)$ breaking effect is about  $0.1\,\rm{GeV}$ \cite{WangEPJC-HL}, then $\mu_{D_{s3}^*}=2.1\,\rm{GeV}$ for the $D_{s3}^*(2860)$, which works well.  Thereafter the  $\overline{MS}$ masses $m_{c}({\rm 2.1\,GeV})$ and $m_s({\rm 2.1\,GeV})$
will be referred to as MS.

The heavy quark masses appearing in  perturbative calculations  are usually taken to be the pole masses.
The $\overline{MS}$ mass $m_c(m_c)$ relates with the pole mass
$m_c$ through the relation \cite{PDG},
\begin{eqnarray}
m_c&=& m_c(m_c)\left[1+\frac{4 \alpha_s(m_c)}{3\pi}+\cdots\right]\, .
\end{eqnarray}
 We can take the approximation $m_c\approx m_c(m_c)$ without the perturbative  corrections for consistency. The value listed
in the Particle Data Group is $m_c(m_c)=(1.275\pm0.025) \,\rm{GeV}$ \cite{PDG}, it is reasonable to take the pole mass
$m_c=(1.275\pm0.025)\,\rm{GeV}$. Up to  corrections of the  order $\mathcal{O}\left({\alpha_s}^3\right)$, the $\overline{MS}$ mass $m_c(m_c)=(1.275\pm0.025) \,\rm{GeV}$ corresponds to the pole mass $m_c=(1.67\pm0.07)\,\rm{GeV}$ \cite{PDG}, which is too large for the QCD sum rules, as $2m_c>M_{J/\psi}={3.096916\,\rm{GeV}}>M_{\eta_c}=2.9836\,\rm{GeV}$ \cite{PDG}.
We can also take the pole masses
$m_c=(1.275\pm0.025)\,\rm{GeV}$ and $m_s=m_s(1{\rm GeV})=0.13\,\rm{GeV}$ \cite{PDG}.

Now we  sum up the  QCD input parameters, which are shown  explicitly in Table 1. There are four sets of input parameters after taking into account the   $\overline{MS}$ masses and pole masses, see Table 2 and Table 3.

\begin{table}
\begin{center}
\begin{tabular}{|c|c|c|c|c|c|c|c|}\hline\hline
Notations & Values                           \\ \hline
SV        & $\langle\frac{\alpha_s GG}{\pi}\rangle=0.012 \,\rm{GeV}^4 $, \, $\langle g_s^3 GGG\rangle=0.045\,\rm{GeV}^6$ \cite{SVZ79,Reinders85,CReview}  \\ \hline
UV        & $\langle \frac{\alpha_s GG}{\pi}\rangle=(0.022 \pm 0.004)\,\rm{GeV}^4 $,\, $\langle g_s^3 GGG\rangle=(0.616\pm0.385)\,\rm{GeV}^6$ \cite{gg-conden} \\ \hline
MS        & $m_{c}({\rm 2.1\,GeV})$, \,  $m_s({\rm 2.1\,GeV})$ \cite{PDG}      \\ \hline
PM        & $m_c=(1.275\pm0.025)\,\rm{GeV}$, \, $m_s=0.13\,\rm{GeV}$ \cite{PDG} \\ \hline
 \hline
\end{tabular}
\end{center}
\caption{ The QCD input parameters, where the SV, UV,  MS and PM denote the standard values, the updated values,  the $\overline{MS}$ masses and the pole masses, respectively,  the $\overline{MS}$ masses $m_{c}({\rm 2.1\,GeV})$ and  $m_s({\rm 2.1\,GeV})$ are obtained according to Eq.(28).  }
\end{table}

The $D_{s3}^*(2860)$ is a conventional meson, we take it for granted that the energy gap between the ground state and the first radial excited state is about $0.5\,\rm{GeV}$.  The measured mass and width are
$M_{D_{s3}^*}=(2860.5\pm 2.6 \pm 2.5\pm 6.0)\,\rm{ MeV}$ and $\Gamma_{D_{s3}^*}=(53 \pm 7 \pm 4 \pm 6)\,\rm{ MeV}$, respectively  \cite{LHCb7574,LHCb7712}. So we take  the threshold parameter $\sqrt{s_0}=2.9+(0.4-0.6)\,\rm{GeV}$
 to avoid the contaminations of the high resonances and continuum states.
We impose the two criteria (pole dominance and convergence of the operator product
expansion) of the QCD sum rules  on the $D_{s3}^*(2860)$, and search for the optimal  values of the Borel parameters. The resulting Borel parameters and pole contributions are shown explicitly in Table 2.  From Table 2, we can see that the pole dominance is well satisfied.

\begin{table}
\begin{center}
\begin{tabular}{|c|c|c|c|c|c|c|c|}\hline\hline
   Input parameters      & $T^2 (\rm{GeV}^2)$   & pole             \\ \hline
   MS+SV                 & $1.9-2.5$            & $(46-78)\%$     \\ \hline
   MS+UV                 & $1.7-2.3$            & $(51-83)\%$     \\ \hline
   PM+SV                 & $1.6-2.2$            & $(51-85)\%$     \\ \hline
   PM+UV                 & $1.4-2.0$            & $(57-90)\%$     \\ \hline
 \hline
\end{tabular}
\end{center}
\caption{ The Borel parameters  and pole contributions of the QCD sum rules. }
\end{table}

In Fig.3, we plot the contributions  come from different terms in the operator product expansion with variations of the  Borel parameters $T^2$. From the figure we can see that if the   $\overline{MS}$ masses are chosen, the contributions of the three-gluon condensate reach zero at about $T^2=0.9\,\rm{GeV}^2$, on the other hand, if the pole masses are chosen, the contributions of the three-gluon condensate reach zero at about $T^2=1.0-1.2\,\rm{GeV}^2$. From Table 2, we can see that the Borel parameters are larger than the low bound  $T^2=0.9\,\rm{GeV}^2$ or $T^2=1.0-1.2\,\rm{GeV}^2$, the operator product expansion is well convergent. In calculations, we observe that it is impossible to obtain the Borel platforms at the low   bound  $T^2=0.9\,\rm{GeV}^2$ or $T^2=1.0-1.2\,\rm{GeV}^2$, and postpone  the Borel parameters to larger values.  The  two criteria of the QCD sum rules are fully satisfied, so we expect to obtain reasonable predictions.

\begin{figure}
\centering
\includegraphics[totalheight=6cm,width=7cm]{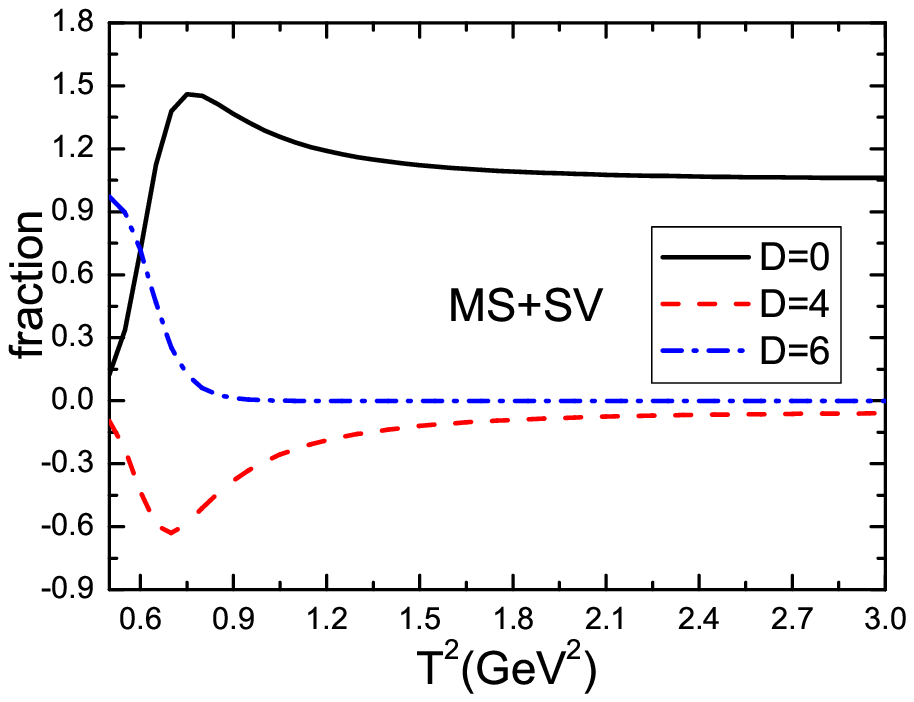}
\includegraphics[totalheight=6cm,width=7cm]{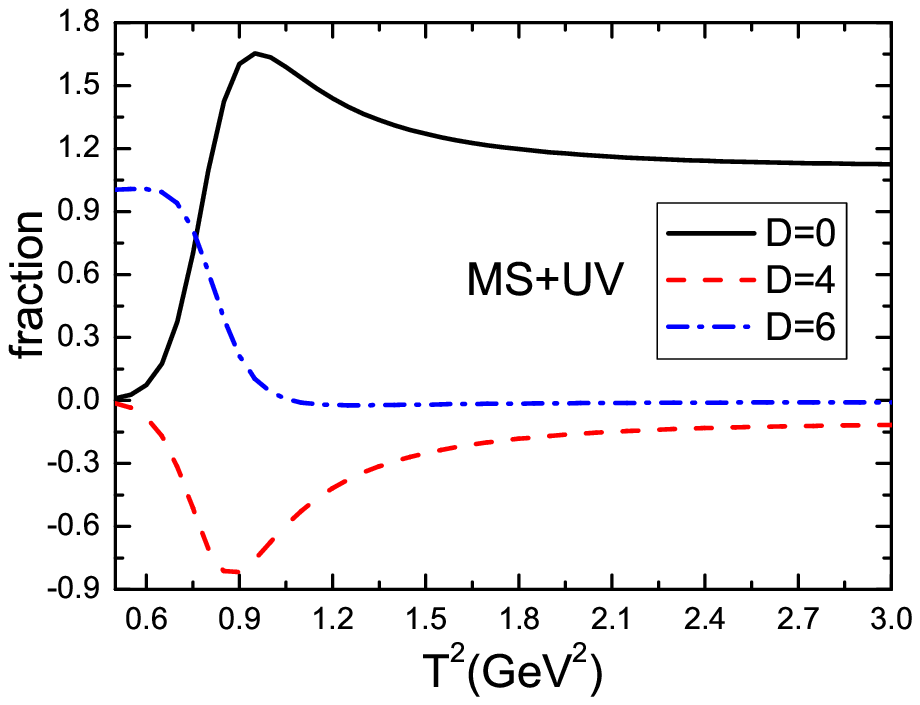}
\includegraphics[totalheight=6cm,width=7cm]{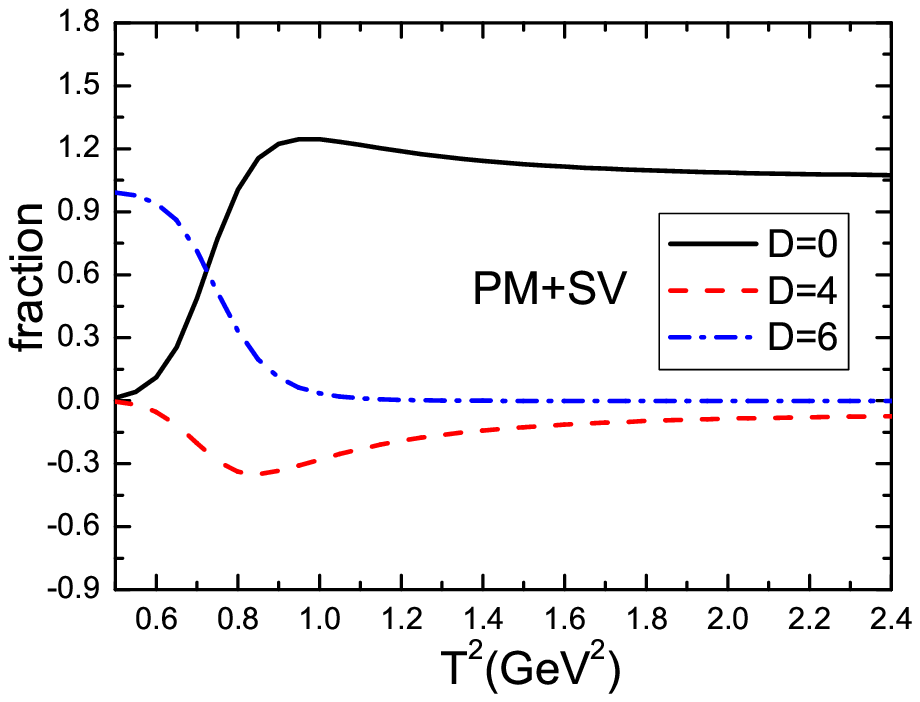}
\includegraphics[totalheight=6cm,width=7cm]{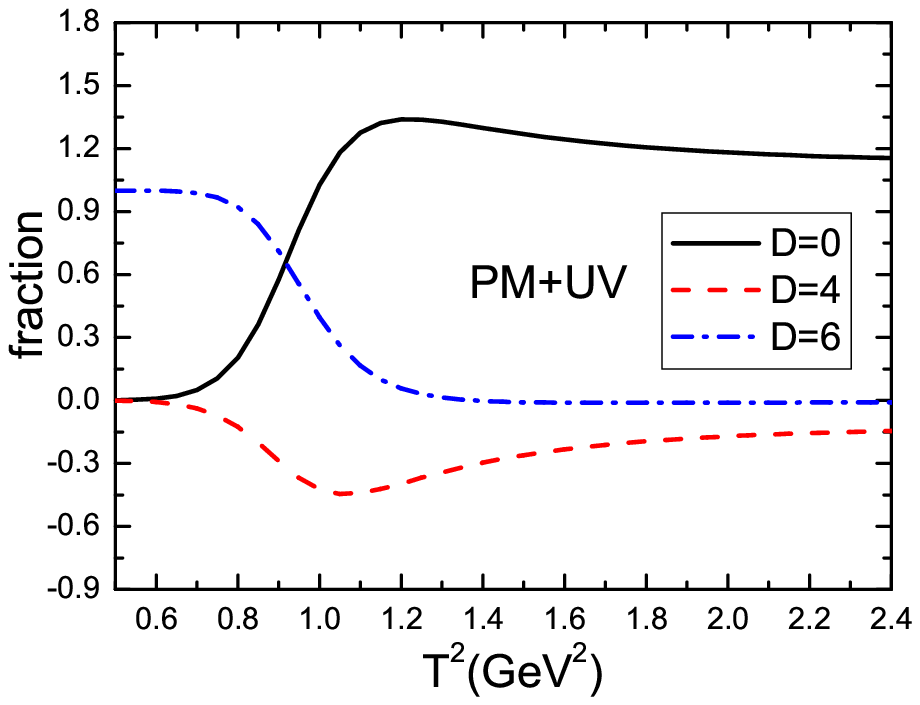}
  \caption{ The  contributions come from different terms in the operator product expansion with variations of the  Borel parameters $T^2$, where $D=0$, $4$ and $6$ denote the dimensions of the vacuum condensates. }
\end{figure}

 Now we take into account the uncertainties of the input parameters and obtain the mass and decay constant of the
 $D_{s3}^*(2860)$, which are shown explicitly in Figs.4-5, and Table 3.  From the figures, we can see that they are rather stable with variations of the Borel parameters  in the Borel windows, it is reliable to extract the mass and decay constant.

 The predicted mass $M_{D_{s3}^*}=(2.86\pm0.10)\,\rm{GeV}$ is in excellent agreement with the experimental value $M_{D_{s3}^*}=(2860.5\pm 2.6 \pm 2.5\pm 6.0)\,\rm{ MeV}$ from the LHCb collaboration  \cite{LHCb7574,LHCb7712}. The calculations based on the QCD sum rules also support assigning the $D_{s3}^*(2860)$ to be the D-wave $\bar{c}s$ meson, and the predicted decay constant $f_{D_{s3}^*}$  can be used to study the hadronic coupling constants involving the $D_{s3}^*(2860)$ with the three-point QCD sum rules or the light-cone QCD sum rules. In the four cases, MS+SV, MS+UV, PM+SV, PM+UV, the predicted mass $M_{D_{s3}^*}=(2.86\pm0.10)\,\rm{GeV}$ remains the same, but the predicted decay constant varies greatly.   It is not unreasonable, as we extract the mass and decay constant from different Borel windows, different Borel windows correspond to different predictions, we choose the special Borel windows to reproduce the experimental value of the mass. The decay constant $f_{D_{s3}^*}$ cannot be extracted from the experimental data, we have to calculate it by some theoretical methods,  the true value cannot be obtained. So in calculating the hadronic coupling constants or form-factors involving the $D_{s3}^*(2860)$ with the three-point QCD sum rules, we must use the value of the  decay constant $f_{D_{s3}^*}$ in a consistent way. The average value is about $f_{D_{s3}^*}=5.46\pm1.02\,\rm{GeV}^4$.

 The measured mass and width are
$M_{D_{s3}^*}=(2860.5\pm 2.6 \pm 2.5\pm 6.0)\,\rm{ MeV}$ and
$\Gamma_{D_{s3}^*}=(53 \pm 7 \pm 4 \pm 6)\,\rm{ MeV}$ \cite{LHCb7574,LHCb7712}, the threshold parameter $\sqrt{s_0}>M_{D_{s3}^*}+\frac{\Gamma_{D_{s3}^*}}{2}\approx 2.9\,\rm{GeV}$.  Now we vary the threshold parameter at a larger interval, $\sqrt{s_0}=3.4\pm 0.3 \,\rm{GeV}$ in stead of $\sqrt{s_0}=3.4\pm0.1\,\rm{GeV}$, then the  uncertainty $\delta \sqrt{s_0}$ leads to the uncertainty $\delta M_{D_{s3}^*}=\pm 0.14\,\rm{GeV}$, $\pm 0.13\,\rm{GeV}$,  $\pm 0.13\,\rm{GeV}$, $\pm 0.12\,\rm{GeV}$ in stead of $\delta M_{D_{s3}^*}=\pm 0.05\,\rm{GeV}$, $0.05\,\rm{GeV}$, $0.05\,\rm{GeV}$, $0.04\,\rm{GeV}$ for the input parameters  MS+SV, MS+UV, PM+SV,
PM+UV, respectively.  The larger uncertainty $\delta\sqrt{s_0}= \pm 0.3 \,\rm{GeV}$ leads to additional uncertainty about $\delta M_{D_{s3}^*}/M_{D_{s3}^*}\approx \pm 3\%$ compared to the uncertainty $\delta\sqrt{s_0}= \pm 0.1 \,\rm{GeV}$.

\begin{figure}
 \centering
 \includegraphics[totalheight=6cm,width=7cm]{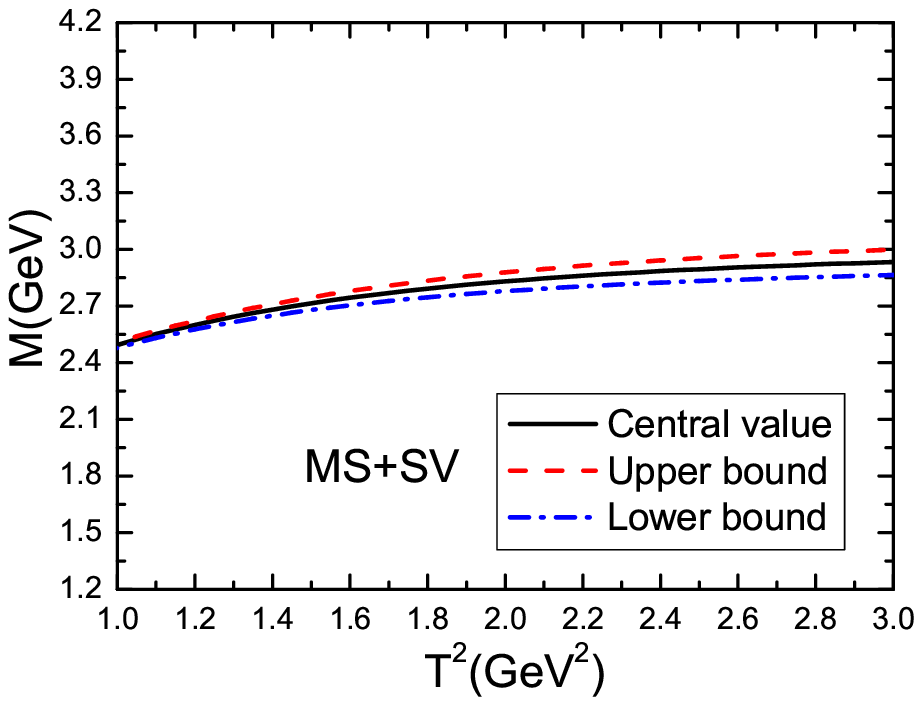}
 \includegraphics[totalheight=6cm,width=7cm]{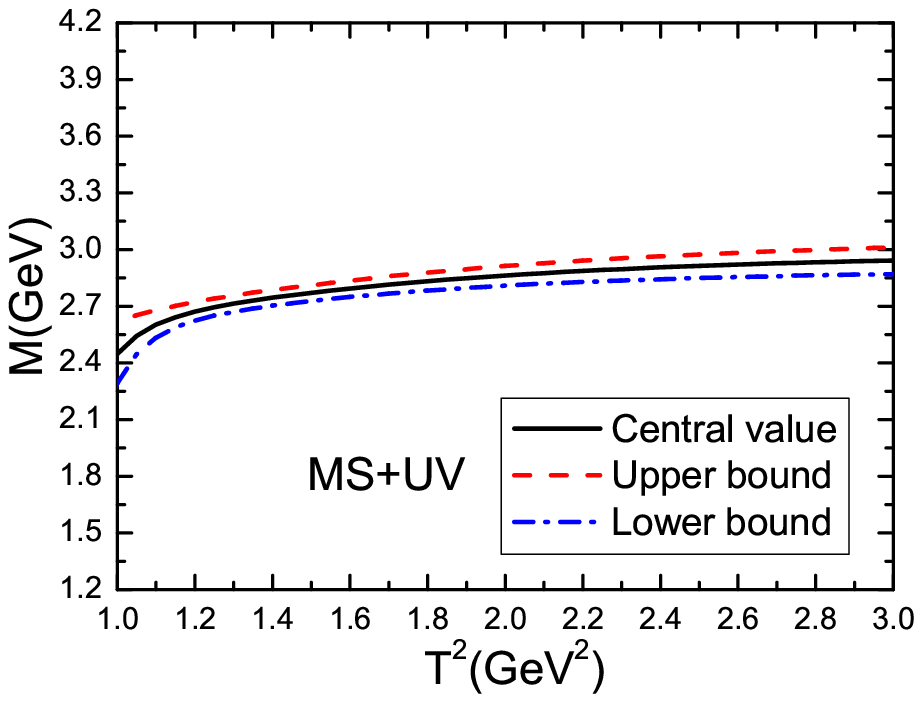}
 \includegraphics[totalheight=6cm,width=7cm]{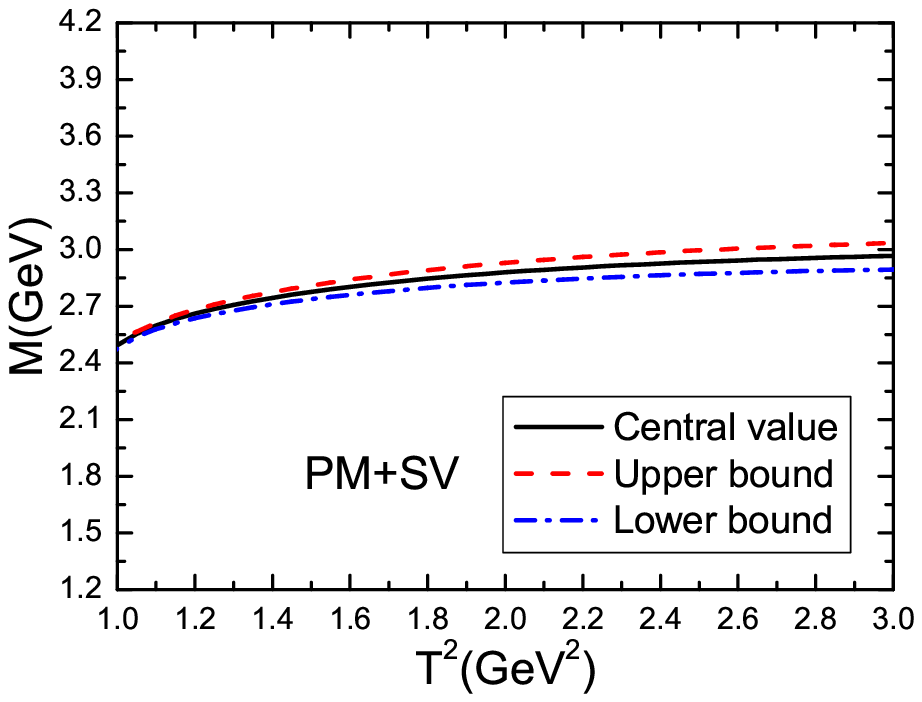}
 \includegraphics[totalheight=6cm,width=7cm]{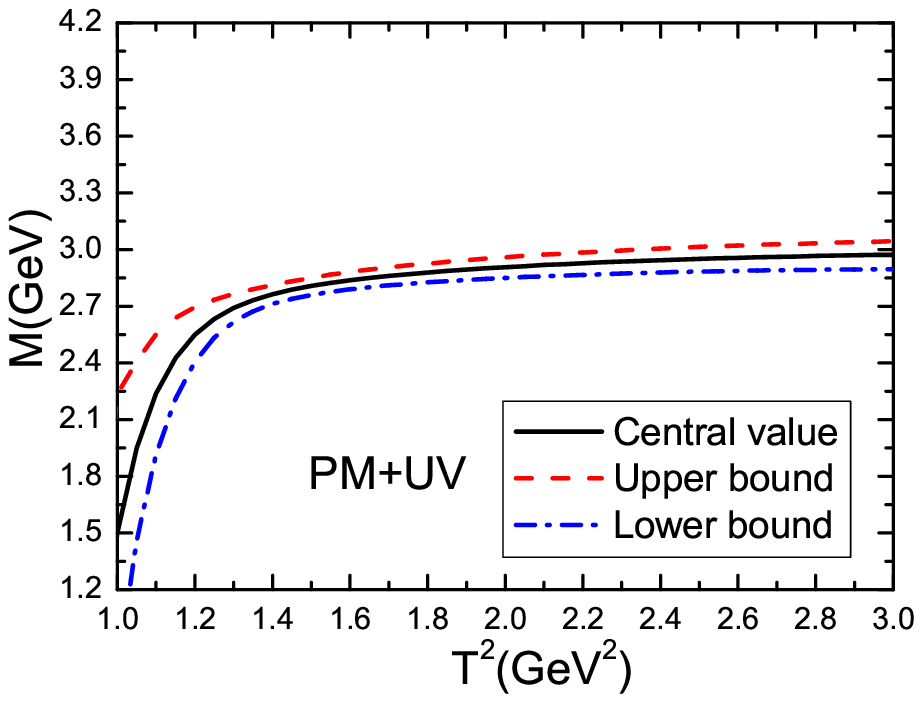}
         \caption{ The mass  $M_{D_{s3}^*}$  with variation of the Borel parameter $T^2$.  }
 \end{figure}

\begin{figure}
 \centering
 \includegraphics[totalheight=6cm,width=7cm]{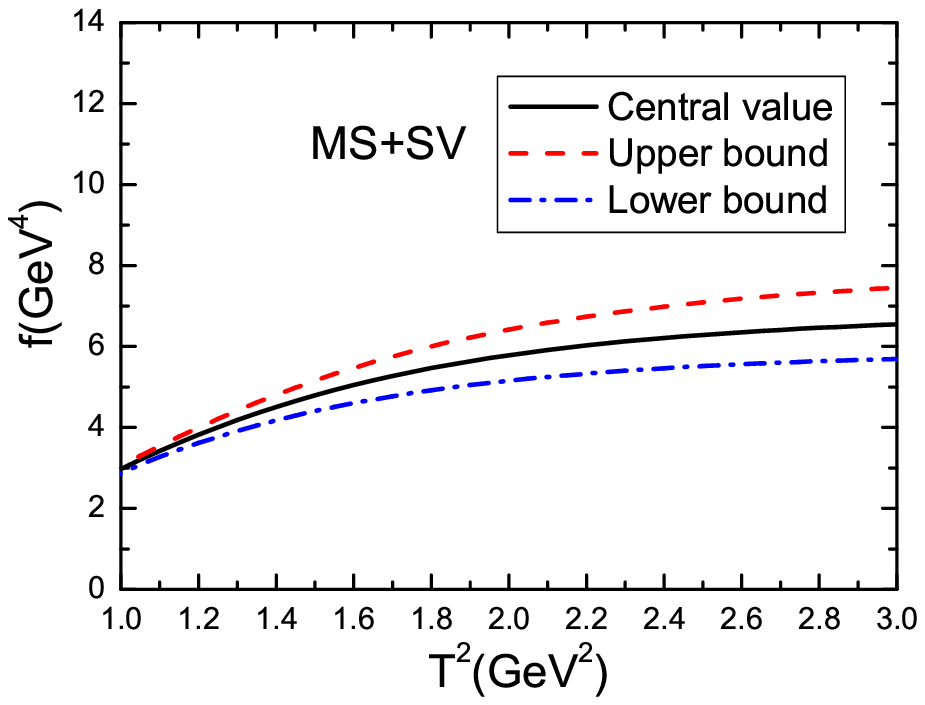}
 \includegraphics[totalheight=6cm,width=7cm]{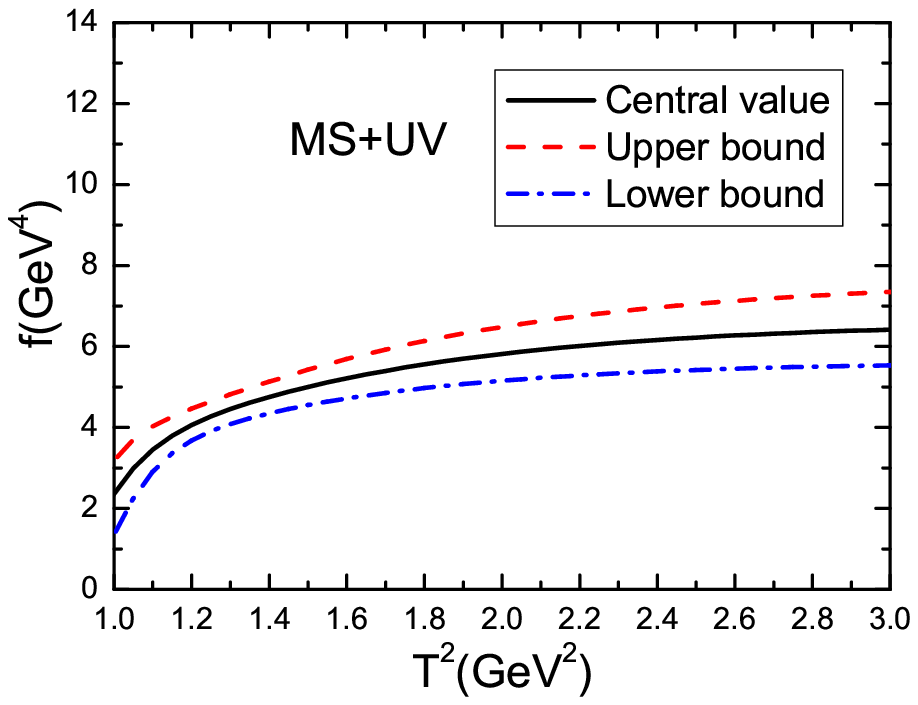}
 \includegraphics[totalheight=6cm,width=7cm]{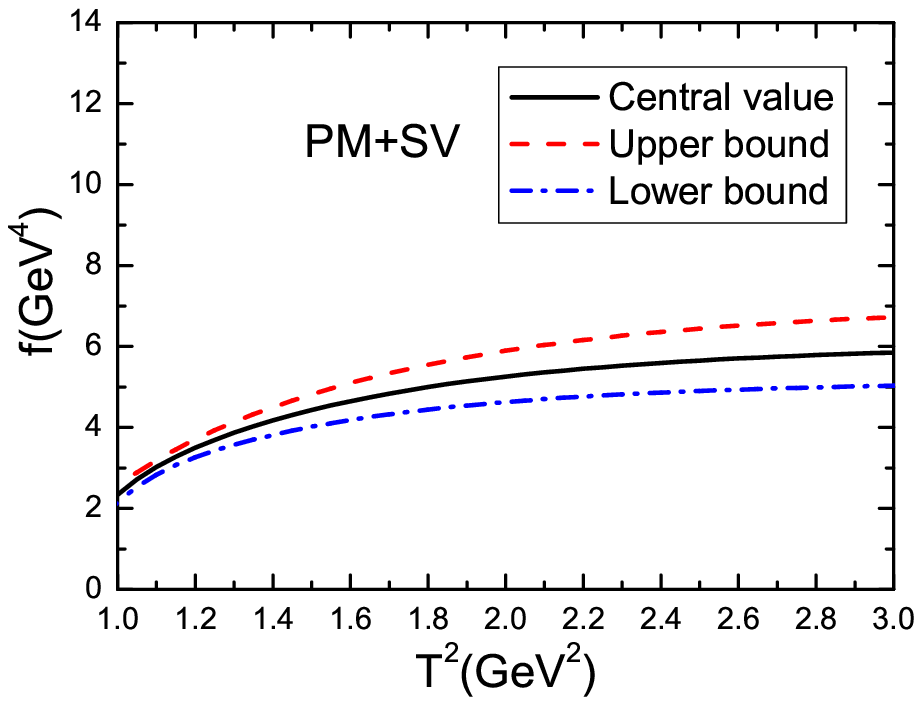}
 \includegraphics[totalheight=6cm,width=7cm]{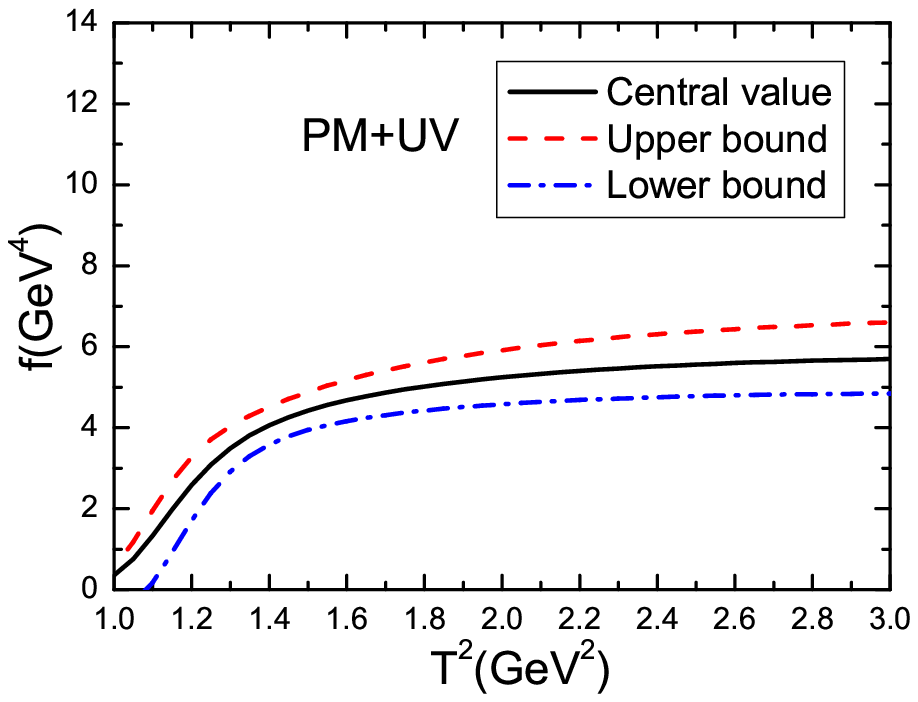}
         \caption{ The decay constant  $f_{D_{s3}^*}$  with variation of the Borel parameter $T^2$.  }
 \end{figure}

Now we explore which interplaiting current is preferred.  In Figs.6-7, we plot the mass and decay constant in the case "MS+SV" with variations of the Borel parameter in the QCD sum rules, where  the currents $J_{\mu\nu\rho}(x)$ and $\eta_{\mu\nu\rho}(x)$ are chosen  or only the perturbative terms are included. From the figures, we can see that the gluon condensate and three-gluon condensate play an important in determining the Borel window even in the case small values of the vacuum condensates are chosen, see Table 1, while in the Borel window, they play a minor important role. Furthermore, we can obtain better QCD sum rules for the current $J_{\mu\nu\rho}(x)$ compared to the current $\eta_{\mu\nu\rho}(x)$, so the covariant   derivative is preferred in constructing the currents.

\begin{figure}
 \centering
 \includegraphics[totalheight=7cm,width=10cm]{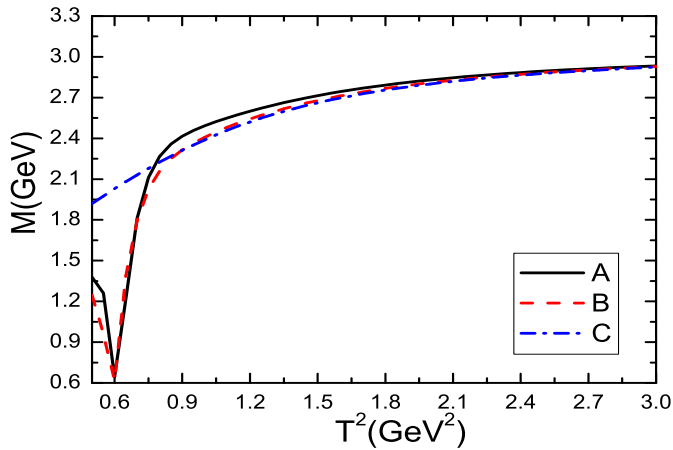}
         \caption{ The mass  $M_{D_{s3}^*}$  with variation of the Borel parameter $T^2$, where the $A$ and $B$ come from QCD sum rules for the currents $J_{\mu\nu\rho}(x)$ and $\eta_{\mu\nu\rho}(x)$, respectively, the $C$ comes from the QCD sum rules where only the perturbative terms are included.  }
 \end{figure}

\begin{figure}
 \centering
 \includegraphics[totalheight=7cm,width=10cm]{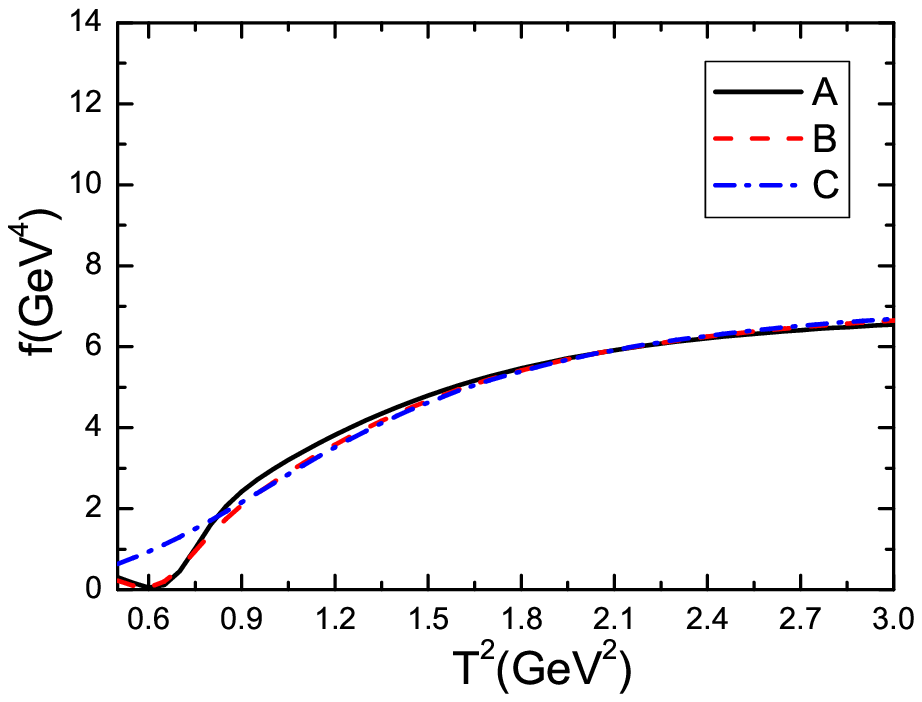}
         \caption{ The decay constant  $f_{D_{s3}^*}$  with variation of the Borel parameter $T^2$, where the $A$ and $B$ come from QCD sum rules for the currents $J_{\mu\nu\rho}(x)$ and $\eta_{\mu\nu\rho}(x)$, respectively, the $C$ comes from the QCD sum rules where only the perturbative terms are included.  }
 \end{figure}

\begin{table}
\begin{center}
\begin{tabular}{|c|c|c|c|c|c|c|c|}\hline\hline
   Input parameters      & $ M_{D_{s3}^*} (\rm{GeV})$   & $f_{D_{s3}^*}(\rm{GeV}^4)$             \\ \hline
   MS+SV                 & $2.86\pm0.10$                & $6.02\pm1.02$     \\ \hline
   MS+UV                 & $2.86\pm0.10$                & $5.82\pm1.01$     \\ \hline
   PM+SV                 & $2.86\pm0.10$                & $5.14\pm1.00$     \\ \hline
   PM+UV                 & $2.86\pm0.10$                & $4.87\pm1.05$     \\ \hline
 \hline
\end{tabular}
\end{center}
\caption{ The mass and decay constant from the QCD sum rules with different input parameters. }
\end{table}

In Fig.8, we plot the predicted mass $M_{D_{s3}^*}$  with variations of the  Borel parameter $T^2$ and $c$-quark mass $m_c$ for both the standard values and updated values of the gluon condensate and three-gluon condensate. From the figure, we can see that all the lineshapes   of the predicted mass $M_{D_{s3}^*}$ cross the experimental value, at the vicinity of the   crossover points, the  lineshapes of the predicted mass $M_{D_{s3}^*}$ with smaller $c$-quark mass are more flat than that with larger $c$-quark mass.  The $\overline{MS}$ mass $m_c(2.1\,\rm{GeV})=1.120\,\rm{GeV}$ is much smaller than the pole mass $m_c=1.275\,\rm{GeV}$, we prefer the $\overline{MS}$ mass. The $c$-quark mass $m_c=1.5\,\rm{GeV}$ is meaningless as
$2m_c>M_{\eta_c}=2.9836\,\rm{GeV}$ \cite{PDG}, the value $m_c=1.5\,\rm{GeV}$ should be discarded.
In calculations, we observe that   at the vicinity of the  crossover point of the lineshapes $M_{D_{s3}^*}=2.86\,\rm{GeV}$ and $m_c=1.1\,\rm{GeV}$,  if the same Borel parameter $T^2$ is chosen, the predicted mass $M_{D_{s3}^*}$ increases about $0.03\,\rm{GeV}$ with the increasement $\delta m_c=0.1\,\rm{GeV}$; while at the vicinity of the  crossover point of the lineshapes $M_{D_{s3}^*}=2.86\,\rm{GeV}$ and $m_c=1.3\,\rm{GeV}$,  if the same Borel parameter $T^2$ is chosen, the predicted mass $M_{D_{s3}^*}$ decreases about $(0.03-0.04)\,\rm{GeV}$ with the decreasement $\delta m_c=-0.1\,\rm{GeV}$.

\begin{figure}
\centering
\includegraphics[totalheight=6cm,width=7cm]{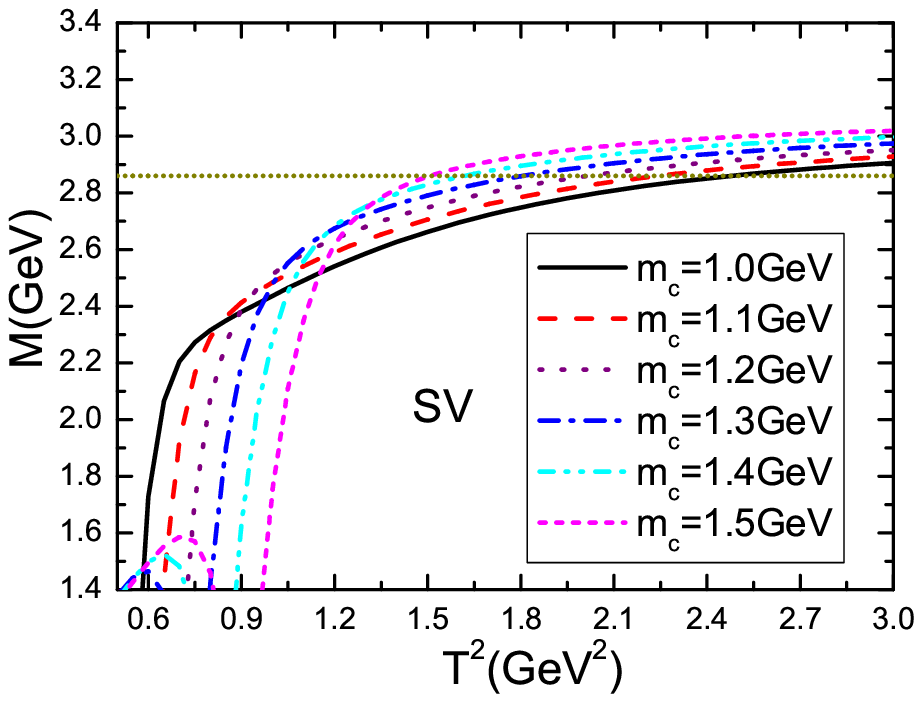}
\includegraphics[totalheight=6cm,width=7cm]{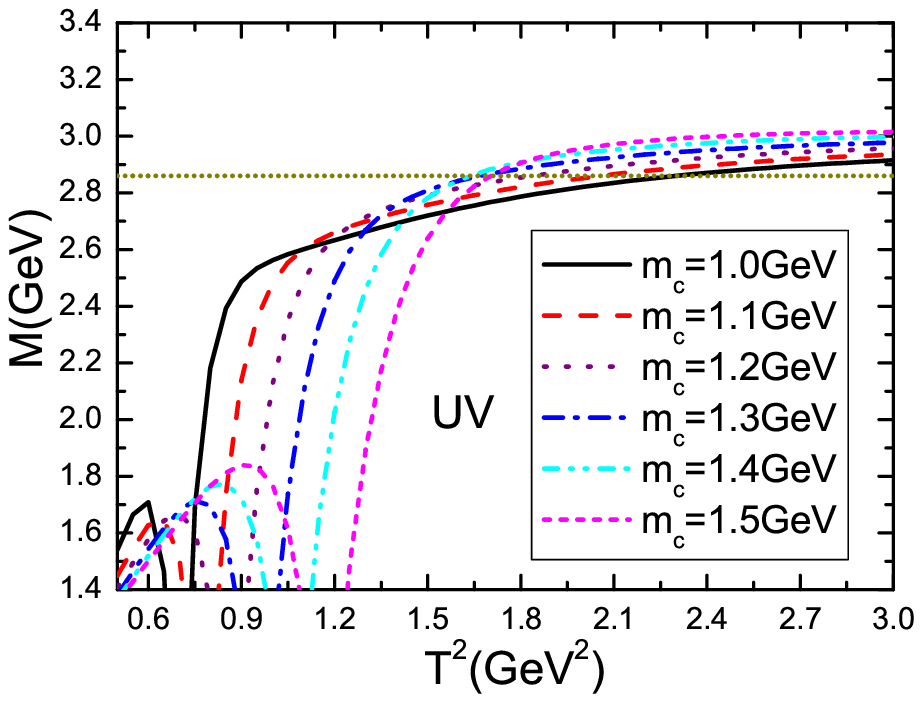}
  \caption{ The  mass $M_{D_{s3}^*}$  with variations of the  Borel parameter $T^2$ and the $c$-quark mass $m_c$, where the horizontal line denotes  the experimental value. }
\end{figure}

\section{Conclusion}
In this article, we assign the $D_{s3}^*(2860)$ to be a D-wave $c\bar{s}$  meson, and  study the mass and decay constant of the $D_{s3}^*(2860)$ with  the QCD sum rules by  calculating the contributions of the vacuum condensates up to dimension-6 in the operator product expansion. The predicted mass $M_{D_{s3}^*}=(2.86\pm0.10)\,\rm{GeV}$ is in excellent agreement with the experimental value $M_{D_{s3}^*}=(2860.5\pm 2.6 \pm 2.5\pm 6.0)\,\rm{ MeV}$ from the LHCb collaboration. The prediction   supports assigning the $D_{s3}^*(2860)$ to be the D-wave $\bar{c}s$ meson. While the predicted decay constant $f_{D_{s3}^*}$  can be used to study the hadronic coupling constants involving the $D_{s3}^*(2860)$ with the three-point QCD sum rules or the light-cone QCD sum rules.

\section*{Acknowledgements}
This  work is supported by National Natural Science Foundation,
Grant Numbers 11375063,  and Natural Science Foundation of Hebei province, Grant Number A2014502017.

\end{document}